# Calculation of material properties for arbitrary shape transformation media

Jin Hu, Xiaoming Zhou and Gengkai Hu*
School of Science, Beijing Institute of Technology, Beijing 100081, P.R.China
*Corresponding author: hugeng@bit.edu.cn

*Abstract*-We propose a general method to evaluate the material parameters for arbitrary shape transformation media. By solving the original coordinates in the transformed region via Laplace's equations, we can obtain the deformation field numerically, in turn the material properties of the devices to be designed such as cloaks, rotators or concentrators with arbitrary shape. Devices which have non-fixed outer boundaries, such as beam guider, can also be designed by the proposed method. Examples with full wave simulation are given for illustration. In the end, wave velocity and energy change in the transformation media are discussed with help of the deformation view.

## I. INTRODUCTION

There has been recently considerable interest in using the coordinate transformation [1,2] to control electromagnetic wave. Many interesting devices have been proposed, such as cloaks [2,3], rotators [4], concentrator [5], parallel beam shifter [6], beam splitter [6] and beam guider [7]. Such devices usually have complex material properties with spatial variation, difficult to be realized by nature material. However, this difficulty can be overcome with the development of metamaterials [3]. Until now, all these devices are limited to simple construction where the analytical transformation function can be found. Recently Yan *et al.* [8] prove that an arbitrary cloak is possible, but how to construct this arbitrary cloak is still lacking. Zharova *et al.* [9] propose a complex shape cloak that the magnetic field inside the cloak is obtained by direct transformation and the cloak shape is limited by a uniform transformation function. Jiang *et al.* [5,10] utilize nonuniform *B*-spline (NURBS) function to represent the geometrical boundary of a concentrator or a cloak, its shape is restricted by the outer boundary which must be self similar to the inner boundary. So the device shapes that they treat with are far from "*arbitrary*". Recently, the authors [11] have proposed a general and flexible method to design cloaks, rotators or concentrators with arbitrary shape based on deformation theory, the outer or inner boundaries of these devices can be of arbitrary shape. In this paper, we will demonstrate this method by designing an *arbitrary* cloak combined with an *arbitrary* rotation-concentrator; we will also extend this method to design *arbitrary* beam guider, which can guide beam to propagate along an *arbitrary* path and in an *arbitrary* direction. In the end, we will discuss the wave velocity and energy change in the transformation media.

## II. CALCULATION METHOD

### A. Theory

For completeness, we will briefly review the design method for arbitrary transformation media proposed by Hu *et al.* [11]. The coordinate transformation method is based on form invariant of Maxwell's equations under transformation. Specifically, Maxwell's equations

$$\nabla \times \mathbf{E} = -\mathbf{u}\partial \mathbf{H}/\partial t, \nabla \times \mathbf{H} = +\boldsymbol{\varepsilon}\partial \mathbf{E}/\partial t, \qquad (1)$$

under a mapping transformation

$$\mathbf{x}' = \mathbf{x}'(\mathbf{x}), \mathbf{E}'(\mathbf{x}') = (\mathbf{A}^{\mathrm{T}})^{-1}\mathbf{E}(\mathbf{x}), \mathbf{H}'(\mathbf{x}') = (\mathbf{A}^{\mathrm{T}})^{-1}\mathbf{H}(\mathbf{x}), \quad (2)$$

retain their form

$$\nabla \times \mathbf{E}' = -\mathbf{u}'\partial \mathbf{H}'/\partial t, \nabla \times \mathbf{H}' = -\boldsymbol{\varepsilon}\partial \mathbf{E}'/\partial t, \qquad (3)$$

with the new material parameters, defined as

$$\boldsymbol{\varepsilon}' = \mathbf{A}\boldsymbol{\varepsilon}\mathbf{A}^{\mathrm{T}}/\det \mathbf{A}, \mathbf{u}' = \mathbf{A}\mathbf{u}\mathbf{A}^{\mathrm{T}}/\det \mathbf{A}, \qquad (4)$$

where $A_{ij} = \partial x'_i/\partial x_j$, characterizing property of the mapping; the superscript T means transposition of a tensor, and $\mathbf{x}$ is the point in a Cartesian space $\Omega$ while $\mathbf{x}'$ is the point in a different Cartesian space $\Omega'$ [12].

So for a given function to control the electromagnetic wave in a specific region $\Omega'$, it can be realized by the map (2) which transforms the known electromagnetic field in a simple region $\Omega$ with the known material properties, so the new material properties in $\Omega'$ can be obtained by (4). Finding $\mathbf{A}$ with the desired function is a crucial point for device design, which at present only limits to simple forms.

The key point of our idea is the new understanding of the transformation tensor $\mathbf{A}$. In the continuum mechanics [13], the tensor $\mathbf{A}$ is called the deformation gradient tensor at $\mathbf{x}$. It is well known in the continuum mechanics community that the transformation $\mathbf{x}' = \mathbf{x}'(\mathbf{x})$ will let a material element to undergo a rigid body rotation (described by a proper orthogonal tensor $\mathbf{R}$) and a pure stretch deformation (described by a positive definite symmetric tensor $\mathbf{V}$). So $\mathbf{A} = \mathbf{V}\mathbf{R}$ [13]. Suppose material parameters in the original space are isotropic and $\lambda_i$ ($i$=1,2,3) are the principal stretches (or $\mathbf{V} = \text{diag}[\lambda_1, \lambda_2, \lambda_3]$) for an infinitesimal element in transformed space, then (4) can be written as (for the simplicity,

we let both the permittivity and permeability in the original space are equal to 1) [11]

$$\boldsymbol{\varepsilon}' = \boldsymbol{\mu}' = \mathrm{diag}[\lambda_1/(\lambda_2\lambda_3), \lambda_2/(\lambda_3\lambda_1), \lambda_3/(\lambda_1\lambda_2)] \quad (5)$$

with help of $\mathbf{V}^2 = \mathbf{A}\mathbf{A}^\mathrm{T}$ and $\det(\mathbf{A}) = \lambda_1\lambda_2\lambda_3$. It is seen that the transformed material parameters depend only on the principal stretches of the deformation; the rigid rotation has no effect on the transformed material parameters. The material parameter tensors and the corresponding element deformation tensor have the same principal directions [11]. Thus, we get an alternative calculation method for the transformed material parameters based on deformation field.

There are two points in the proposed method for arbitrary shape transformation media [11]: Firstly, we introduce harmonic displacement field to obtain deformation filed, thus the deformation field is smooth enough to make no reflection inside the region, and the harmonic function can be completely evaluated from the boundary condition [14]. Secondly, we solve the Laplace's equations on the transformed region instead on the original one, thus the boundary condition can be written out for these arbitrary shape boundaries. The Laplace's equations are

$$\Delta_{\mathbf{x}'} x_i = 0, i = 1, 2, 3, \quad (6)$$

and the boundary conditions are

$$\mathbf{x} = f(\mathbf{x}'), \quad \mathbf{x}' \in \partial\Omega'. \quad (7)$$

where $f(\mathbf{x}')$ is the original coordinates of the boundary which was transformed to the boundary $\partial\Omega'$, and $f(\mathbf{x}')$ can be obtained easily form the transformation. Once the Laplace's equations (6) with the boundary conditions (7) are solved, the deformation field can then be evaluated, so as to the transformed material parameters. It is worth to point out that in some cases where the stretches of transformation can be found easily, we can use (5) to calculate transformed material parameters directly. In the following, some examples will be given to illustrate the above ideas.

*B. Examples*

The first example is to design an arbitrary 2D cloak combined with an arbitrary rotation-concentrator. Fig.1 shows the construction of this device. The whole outer boundary is kept fixed. For the cloak, the Dirichlet boundary conditions for the 2D Laplace's equations are

$$\begin{aligned}x_i &= x'_i, & \mathbf{x}' &\in \partial\Omega'_+, \\ x_i &= c_i, & \mathbf{x}' &\in \partial\Omega'_{1-}, \\ i &= 1,2.\end{aligned} \quad (8)$$

where $(c_1, c_2)$ are Cartesian coordinates of a given point which was transformed to the inner boundary $\partial\Omega'_{1-}$. The rotation-concentrator is formed by compressing an arbitrary sub-area inside the arbitrary region in the self-similar way and further rotating the sub-area by a given angle $\theta$. The inner boundary conditions for the arbitrary 2D rotation-concentrator are

$$\begin{aligned}x_1 &= kx'_1\cos\theta + kx'_2\sin\theta, \\ x_2 &= -kx'_1\sin\theta + kx'_2\cos\theta, & \mathbf{x}' &\in \partial\Omega'_{2-},\end{aligned} \quad (9)$$

where $k \neq 0$ is the scale of compression. Here we set both the compression center and rotation center to be (0, 0). The COMSOL Multiphysics finite element-based solver is used for the simulation. Our model includes two PDE modes (Laplace's equations) and an in-plane wave mode (TE). The PDE modes are used to solve the Laplace's equations for the two corresponding original coordinates on the transformed region, and then the deformation can be obtained; the deformation is then used to calculate the transformed material parameters. The in-plane wave mode is used to determine the electromagnetic fields with the determined material parameters. Fig.1 also shows the computational domain, the outer layers of the computational domain are PMLs to mimic the infinite space. Fig.2 shows the resulting simulated electric-field distribution for the device under transverse-electric (TE) polarized plane wave, with $k=2$ and $\theta = \pi/4$. Clearly, cloaking, concentrating and rotating effect are indeed found. The necessary material parameters of the device can be retrieved from the calculation.

The following sample is related to beam guider. A beam guider is used to direct a beam along a desired path and direction; it can be constructed by curving a cube [7]. Thus, its outer boundary is no longer fixed as that of cloaks, rotators or concentrator. To get the boundary condition for an arbitrary beam guider, we classify the beam guider into two types: beam rotator [7] and parallel beam shifter [6], respectively. The former only changes the wavefront and the latter only changes the beam propagation path. So general beam guider can be constructed by assembling beam rotators and parallel beam shifters, as the idea proposed in [7].

The second example is a 2D beam rotator, which is constructed by curving a square cylinder into a sector cylinder. Fig.3 shows the cross section of this transformation. It is easy to show that by this transformation the length of line segment $y = c$ ($0 \leq c \leq a$) changes from side length $a$ to arc length $r\pi/2$, where $r = y$ is the radial coordinate of the sector in polar coordinate system; line elements directed along $\hat{\mathbf{r}}$ and $\hat{\mathbf{z}}$ are keep unchanged. Therefore, the stretches in the cylindrical coordinate system are $\lambda_\theta = (r\pi)/(2a)$, $\lambda_r = \lambda_z = 1$, respectively. Using (5), the transformed materials parameters are very simple:

$$\begin{aligned}\varepsilon'_{\theta'} &= \mu'_{\theta'} = \lambda_\theta/(\lambda_z\lambda_r) = (r\pi)/(2a), \\ \varepsilon'_{r'} &= \mu'_{r'} = \lambda_r/(\lambda_\theta\lambda_z) = (2a)/(r\pi), \\ \varepsilon'_{z'} &= \mu'_{z'} = \lambda_z/(\lambda_r\lambda_\theta) = (2a)/(r\pi).\end{aligned} \quad (10)$$

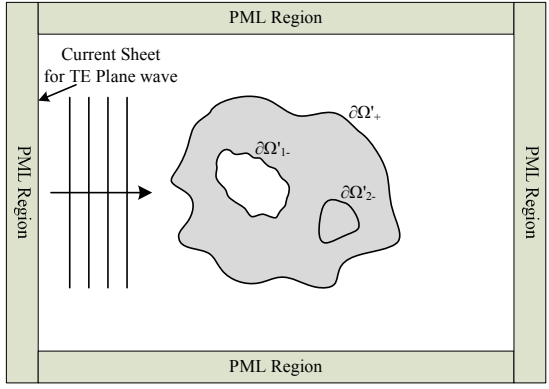

Fig. 1. (Color online) Computational domain for the full-wave simulation of a cloak combined with a rotation-concentrator.

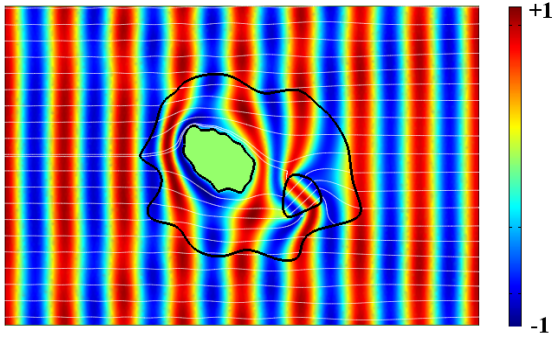

Fig. 2. (Color online) The resulting electric-field distribution near the cloak combined with rotation-concentrator. Power-flow lines (in white) show electromagnetic power.

The material parameters can be evaluated analytically without explicit expression for the transformation. The COMSOL Multiphysics simulation result is show in Fig.4. The incident beam is directed to a new direction as desired. Of course, choosing different central angles for the sector, beam can be guided to different directions. This sample shows that use of (5) can bring considerable convenience in some cases.

The third example is a 2D parallel beam shifter, which can change the beam propagating path from straight line to an arbitrary plane curve, however the final beam propagating direction is kept unchanged. A simple strategy to construct the parallel beam shifter is following: the incident side is kept fixed, and all other sides have translations to follow the desired curve path. Fig.5 shows the transformation of rectangular area to a curved zonary area. So the boundary conditions become

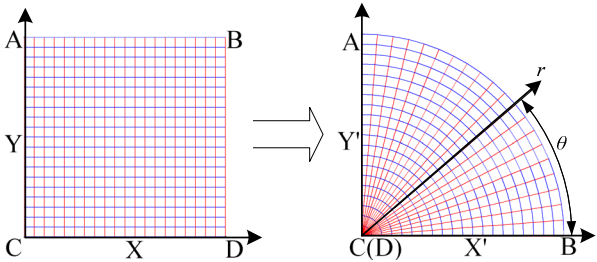

Fig. 3. (Color online) Transformation from original square area to sector area.

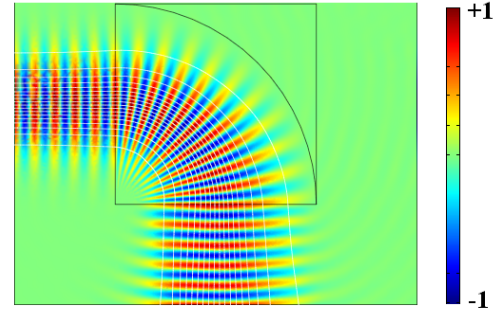

Fig. 4. (Color online) The resulting electric-field distribution of Gaussian beam (TE polarized) in the sector. Beam incident from the left side. Power-flow lines (in white) show electromagnetic power.

$$\begin{aligned} x=x', y=y', & (x',y')\in \partial\Omega'_1, \\ x=x', y=y'-c, & (x',y')\in \partial\Omega'_2, \\ x=x', y=d_1, & (x',y')\in \partial\Omega'_3, \\ x=x', y=d_2, & (x',y')\in \partial\Omega'_4, \end{aligned} \quad (11)$$

where $c$, $d_1$, $d_2$ are constants, as shown in Fig.5. The COMSOL Multiphysics simulation result is shown in Fig.6. Since there is no coordinate transformation in $X$ direction, the incident beam propagates along $X$ axis will keep the same propagating direction when leaving the beam shifter, however, the beam propagation path is indeed redirected inside the transformed region as desired.

If change of wavefront is also needed, we can simply assemble the above mentioned beam rotator on the outcome side $\partial\Omega'_2$. The idea of assembling different elements to obtain more complex functions can bring more flexibility in device design [7]. Fig.7 shows the simulation result for the assemblage. Clearly, both the curved propagation path and the changed wavefront have been achieved.

### III. DISCUSSION AND CONCLUSION

The deformation view on material parameters and fields can also help to understand the inherent characteristics of the transformation method. From (5), it is easy to show that

$$1/\sqrt{\varepsilon_i' \mu_j'} = \lambda_k, i,j,k=1,2,3, i\neq j, i\neq k, j\neq k \quad (12)$$

The left hand sides of (12) are the principal wave velocity in an anisotropic homogeneous medium [15], so $1/\sqrt{\varepsilon_i'\mu_j'}$ can be considered as the change ratio of wave velocity in $\hat{\mathbf{k}}$ direction due to the transformation. Eqs.(12) shows that the change ratio of wave velocity in each principal direction is equal to the stretch in that direction. Therefore, the wave propagation in an element will take the same time before and after the transformation. This result agrees with the fact that the transformations only imposes on the spatial coordinate and not on time [2,12].

Since $(\mathbf{A}^T)^{-1} = [(\mathbf{VR})^T]^{-1} = \mathbf{V}^{-1}\mathbf{R}$, from (2), the transformed electromagnetic fields can be expressed as $\mathbf{E}' = \mathbf{V}^{-1}\mathbf{R}\mathbf{E}$ and $\mathbf{H}' = \mathbf{V}^{-1}\mathbf{R}\mathbf{H}$, respectively. Noting $|\mathbf{R}\mathbf{E}| = |\mathbf{E}|$ and (5), we have $\mathbf{\varepsilon}'\mathbf{E}'\mathbf{E}' = |\mathbf{E}|^2/(\lambda_1\lambda_2\lambda_3)$. Similarly, $\mathbf{\mu}'\mathbf{H}'\mathbf{H}' = |\mathbf{H}|^2/(\lambda_1\lambda_2\lambda_3)$. Therefore the electromagnetic energy

$$\int_{\Omega'} W' dv' = \int_{\Omega} W'(\lambda_1\lambda_2\lambda_3 dv) = \int_{\Omega} W dv \qquad (13)$$

is unchanged before and after the transformation, where $W' = (1/2)(\mathbf{\varepsilon}'\mathbf{E}'\mathbf{E}' + \mathbf{\mu}'\mathbf{H}'\mathbf{H}')$ and $W = (1/2)(\mathbf{E}\mathbf{E} + \mathbf{H}\mathbf{H})$ are electromagnetic energy density in $\Omega'$ and $\Omega$, respectively; $dv$ and $dv'$ are element volumes in $\Omega$ and $\Omega'$, respectively. Eqs.(13) conforms that the energy is conserved during the transformation.

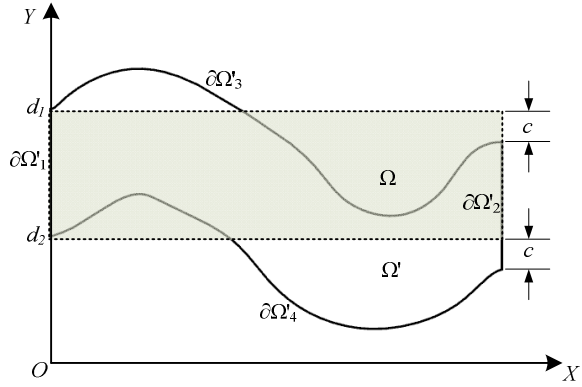

Fig. 5. (Color online) Transformation from original rectangular area to curved zonary area. Rectangular region surrounded by broken line is the original area and region surrounded by solid line is the transformed one.

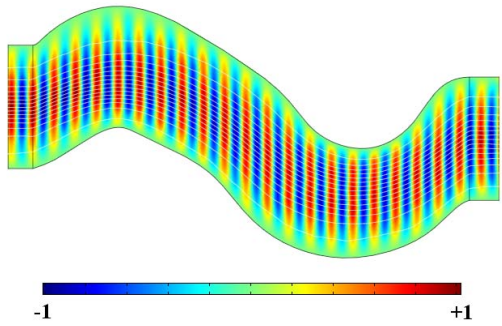

Fig. 6. (Color online) The resulting electric-field distribution of a Gaussian beam (TE polarized) in the parallel beam shifter. Beam incidents form left side. Power-flow lines (in white) show electromagnetic power.

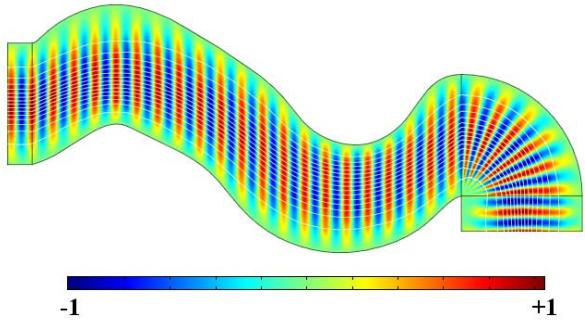

Fig. 7. (Color online) The resulting electric-field distribution of a Gaussian beam (TE polarized) in the assemblage of a parallel beam shifter and a beam rotator. Beam incidents form left side. Power-flow lines (in white) show electromagnetic power.

To conclude, we therefore proposed a flexible method for arbitrary transformation media based on the deformation theory in continuum mechanics. The functionality of a design device is converted to the boundary condition and the transformed material parameters are related to the principle stretches of deformation. The Laplace's equations are proposed to evaluate numerically the deformation field, which are further used to compute the transformed material parameters inside of the cloak. If the stretches of the transformation can be obtained directly, the material parameters calculation becomes very simple. Three examples (an arbitrary 2D cloak combined with an arbitrary rotation-concentrator, beam rotator and parallel beam shifter) are examined to illustrate the feasibility of the method. The full-wave simulations based on Maxwell's equations together with calculations of deformation field are well integrated in a two-step model with help of the finite-element software COMSOL Multiphysics. The simulation results validate the method. The deformation view is also used to find wave velocity and energy change in the transformation.